\documentclass[11pt]{article}
\usepackage{mysetup}
\usepackage{verbatim}
\usepackage[colorlinks,pagebackref,linkcolor=blue, filecolor = blue,
citecolor = blue, urlcolor = blue]{hyperref}

\newcommand{\univ}{{{U}}}
\newcommand{\rngspc}{{\mathcal{R}}}
\newcommand{\ku}[1]{{#1}^{\cup k}}
\newcommand{\kuv}[2]{{#1}^{\cup #2}}
\newcommand{\opt}{OPT}
\newcommand{\loc}{LOC}
\newcommand{\problem}[1]{{\sc #1}\xspace}
\newcommand{\mcp}{\problem{Maximum Coverage}}
\newcommand{\cmcp}{\problem{Constrained Maximum Coverage}}
\newcommand{\mvcp}{\problem{Maximum Vertex Coverage}}
\newcommand{\pvcp}{\problem{Partial Vertex Cover}}
\newcommand{\mhcp}{\problem{Maximum Halfspace Coverage}}
\newcommand{\mrcp}{\problem{Maximum Rectangle Coverage}}
\newcommand{\hlf}{\mathbf{h}}
\newcommand{\rct}{\mathbf{r}}
\newcommand{\poly}{\operatorname{poly}}
\newcommand{\initialAssoc}{A^{\mathrm{init}}}
\newcommand{\newAssoc}{A^{\mathrm{new}}}

\usepackage{amssymb}
\usepackage{mathrsfs}
\usepackage{amsmath}
\usepackage{paralist}

\newcommand{\ignore}[1]{}

\title{Approximating Low-Dimensional Coverage Problems}
\author{Ashwinkumar Badanidiyuru\thanks{Department of Computer Science, Cornell University, Ithaca, NY.  {\tt ashwinkumarbv@gmail.com}.  Supported by NSF grants AF-0910940 and IIS-0905467.} 
\and 
Robert Kleinberg\thanks{Department of Computer Science, Cornell University, Ithaca, NY.  {\tt rdk@cs.cornell.edu}.  Supported by NSF grants AF-0910940, CCF-0643934, and IIS-0905467, AFOSR grant FA9550-09-1-0100, a Microsoft Research
New Faculty Fellowship, a Google Research Grant, and 
an Alfred P. Sloan Foundation Fellowship.}
\and
Hooyeon Lee\thanks{Department of Computer Science, Stanford University, Stanford, CA.}
}
\date{}

\begin{document}
\maketitle
\thispagestyle{empty}

\begin{abstract}
We study the complexity of the maximum coverage problem, restricted to set systems of bounded VC-dimension.  Our main result is a fixed-parameter tractable approximation scheme: an algorithm that outputs a $(1-\eps)$-approximation to the maximum-cardinality union of $k$ sets, in running time $O(f(\eps,k,d)\cdot poly(n))$ where $n$ is the problem size, $d$ is the VC-dimension of the set system, and $f(\eps,k,d)$ is exponential in $(kd/\eps)^c$ for some constant $c$.  We complement this positive result by showing that the function $f(\eps,k,d)$ in the running-time bound cannot be replaced by a function depending only on $(\eps,d)$ or on $(k,d)$, under standard complexity assumptions.

We also present an improved upper bound on the
approximation ratio of the greedy algorithm in special cases 
of the problem,
including when the sets have bounded cardinality and when they are
two-dimensional halfspaces.  Complementing these positive results,
we show that when the sets are four-dimensional halfspaces neither the
greedy algorithm nor local search is capable of improving the 
worst-case approximation ratio of $1-1/e$ 
that the greedy algorithm achieves on 
arbitrary instances of maximum coverage.

\end{abstract}

\newpage
\setcounter{page}{1}

\shortverfalse

\section{Introduction} \svlabel{sec:intro}

The \mcp problem is one of the classical NP-hard
combinatorial optimization problems.  An instance
of \mcp is specified by a triple $(\univ,\rngspc,k)$
where $\univ$ is a finite set, $\rngspc$ is a collection
of subsets of $\univ$, and $k$ is a positive integer.
The objective is to output a $k$-tuple of elements
of $\rngspc$ such that their union contains as many
elements as possible.  (In the \emph{weighted} version
of the problem, elements $x \in \univ$ have non-negative
weights $w(x)$ and the objective is to maximize the
combined weight of elements in the union.)  A very natural
greedy algorithm for \mcp chooses $k$ sets sequentially,
where each new set is chosen to maximize the number (or
combined weight) of elements covered by the new set but
not by any of the preceding ones.  It has been known for
decades that this algorithm has an approximation factor
of $\left( 1 - \frac1e \right)$~\cite{CNW80}; in fact,
it is a special case of the greedy algorithm for
maximizing a monotone submodular function subject to
a cardinality constraint~\cite{NWF}, and the algorithm's
approximation factor remains $\left(1-\frac{1}{e}\right)$
even in this more general case.  It was shown by 
Feige~\cite{FeigeSetCov} that
this approximation factor is the best possible,
even for the unweighted \mcp problem, unless
P=NP.

One of the reasons the greedy algorithm for
\mcp is so widely studied
is that it has innumerable applications: originally
introduced by Cornuejols, Fisher, and Nemhauser~\cite{CFN77} to
model the problem of locating
bank accounts in multiple cities to maximize ``float,''
it was subsequently applied in databases~\cite{HRU-sigmod},
social networks~\cite{KKT03}, sensor placement~\cite{krause08thesis},
information retrieval~\cite{Radlinski/etal/08a,Yue/Joachims/08a},
and numerous other areas.
A prototypical application arises in the information
retrieval setting, when considering the problem of
assembling a list of $k$ documents to satisfy the
information needs of as many users as possible.
Equating every document with the set of users
whom it satisfies, we see that this information
retrieval problem is modeled by the \mcp problem.

Given the extremely broad applicability of \mcp
problems, it is natural
to wonder whether the approximation ratio of $1 - \frac1e$
is the strongest theoretical guarantee one can hope 
for.  Feige's hardness result eliminates the possibility
of obtaining a better worst-case approximation ratio
in polynomial-time, but the problem instances arising
in applications are unlikely to resemble worst-case
instances of \mcp.  Is it possible to identify broad
classes of \mcp instances (hopefully resembling those
that arise in practice) such that the greedy algorithm
provably achieves an approximation factor better than
$1-\frac1e$ on these instances?  If not, can one 
design a different polynomial-time algorithm with an
improved approximation factor?  These are the questions
that inspired our paper.

Let us reconsider the problem of assembling a top-$k$
list of documents, mentioned above, in light of these 
questions.  At least two 
aspects of this application distinguish  it
from an arbitrary instance of \mcp.
\begin{enumerate}[(1)]
\item  The value $k$ is quite small compared to $n$,
the input size.  A typical instance might involve
processing a list of thousands or millions of documents
to extract a list of $k=10$ top choices.
\item  The set system $(\univ,\rngspc)$ is likely to
have a ``low-dimensional'' structure.  For example,
a natural model of users' preferences might assume
that there are $d \ll n$ topics, a document contains
a mix of topics described by a vector in $\R_+^d$, and
the user's information need is satisfied if the
dot product of this vector with another vector
in $\R_+^d$ (describing the mix of topics the user 
seeks to read about) exceeds some threshold.
\end{enumerate}
Is the approximation ratio of the greedy algorithm better than
$1 - \frac1e$ under these circumstances?  If not, is there 
some other algorithm that is significantly better?

We answer the first question negatively and the second
one affirmatively.  More precisely, for $d \geq 4$ we show
that the greedy algorithm's approximation ratio in this
special case is no better than its worst-case approximation
ratio of $1 - \left( 1 - \frac1k \right)^k \approx 1 - \frac{1}{e}$,
but that there is an algorithm with running time 
$O( f(\eps,k,d)  \cdot poly(n) )$ whose approximation factor
is $(1-\eps)$, for some function $f(\eps,k,d)$.
(Of course, for \emph{very} small values of $k$ a
trivial brute-force search over all collections of
$k$ sets in $\rngspc$ finds an exactly optimal solution
in $O(n^{k+1})$ time, but a fixed-parameter algorithm
whose running time is exponential in $k$ but 
quadratic in $n$ is vastly faster when $k=10$ and
$n=10^6$, for instance.)

The following subsection describes our contributions
in more detail.

\subsection{Our contributions}

Our main contribution is a fixed-parameter approximation
scheme (fpt-AS) for the \mcp problem, parameterized by the
number of sets $k$, the approximation parameter $\eps$,
and the VC-dimension of the set system, $d$.  
Letting $n$ denote the problem size --- i.e.\ the sum
of cardinalities of all the sets in $\rngspc$ --- the
approximation scheme has running time $O( f(\eps,k,d) \cdot poly(n))$,
where $f(\eps,k,d) = \exp \left( \tilde{O}(k^2 d \eps^{-5}) \right)$\footnote{$\tilde{O}$ hiding log factors}.
The algorithm, which is presented in Section~\svref{sec:fpt-as},
is based on three ingredients.  First, set systems of bounded
VC-dimension have bounded-size $\eps$-approximations (see
Section~\svref{sec:fpt-as} for definitions) and there is
even a deterministic algorithm to find them in linear time~\cite{ChazMat}.
Second, this means it is easy to design a fpt-AS for the special case of \mcp
in which the set system has bounded VC-dimension \emph{and}
the optimum solution covers a constant fraction of
the elements.  Third, the general case can be reduced
to this special case by an intricate non-deterministic
algorithm, which can then be made deterministic at the
cost of blowing up the running time by a factor that is
exponential in $k^2 d \eps^{-5}$, but independent of $n$.

In Section~\svref{sec:lb} we show that various aspects
of this result cannot be improved, under standard complexity
assumptions.  First, the function $f(\eps,k,d)$ cannot be
replaced by a function depending polynomially on $k$ unless
$P=NP$.  Second, it cannot be replaced by a function
depending polynomially on $1/\eps$ unless $P = W[1]$.
(The question of whether the exponential dependence on $d$ can be 
eliminated is intriguing, but it is unlikely to be easily
resolvable since fixed-parameter complexity theory lacks
machinery analogous to the PCP Theorem 
for proving $W[1]$-hardness of approximation.)
Furthermore, these hardness results apply even 
in some very simple cases: \mcp with set systems
of VC-dimension 2, or with halfspaces in dimension 4,
or with rectangles in dimension 2.  Moreover, in
all three of these special cases, the greedy
algorithm fails to achieve an approximation factor
better than $1 - \frac1e$.

These negative results about the greedy algorithm
are counterbalanced by some positive results
that we present in Section~\svref{sec:boundedCardinality}.
We identify a parameter of the problem instance --- the
\emph{covering multiplicity}, denoted by $r$ ---
such that the greedy algorithm's approximation factor 
is never worse than $1 - (1 - \frac{1}{r})^r$.
The covering multiplicity satisfies $r \leq k$, and
when the inequality is strict this improves upon
the worst-case approximation bound for the greedy algorithm.
For problem instances whose sets have cardinality at most $r$,
the covering multiplicity is bounded by $r$, and for instances
in which the sets are two-dimensional halfspaces the covering
multiplicity is bounded by 2, implying that the greedy
algorithm is a $\frac34$-approximation in the latter case.

\subsection{Related work}

As mentioned above, the \mcp problem was introduced,
and the greedy algorithm analyzed, by
Conuejols et al.\ in~\cite{CFN77}.  This work
was subsequently generalized to the context of
submodular functions by Nemhauser et al.~\cite{NWF}.
A matching hardness of approximation for \mcp was
obtained by Feige~\cite{FeigeSetCov}, in a paper 
that also settled the approximation hardness of
the closely related \problem{Set Cover} problem,
establishing that the greedy algorithm achieves
the optimal approximation ratio (up to lower order
terms) for both problems.

The approximability of special cases of 
\problem{Set Cover} and \mcp was subsequently
investigated in numerous papers.  For example,
the \mvcp problem is the special case of \mcp
in which $\univ$ is the edge set of a graph and
every set in $\rngspc$ is the set of edges 
incident to one vertex of that graph.  This
special case of the problem was shown to be
APX-hard by Petrank~\cite{Petrank}.
A landmark
paper by Ageev and Sviridenko~\cite{AgeevSviri}
introduced the technique of \emph{pipage rounding}
and used it to give a (non-greedy) 
polynomial-time $\frac34$-approximation
algorithm for \mvcp; more generally, they
gave a polynomial-time algorithm 
with approximation factor 
$\left( 1 - \left(1 - \frac{1}{k} \right)^k \right)$
for the special case in which every element of
$\univ$ belongs to at most $k$ sets in $\rngspc$.

Computational geometers have intensively studied
special cases of \problem{Set Cover}, or the dual
problem of \problem{Hitting Set}, when
the set system is defined geometrically,
e.g.\ by rectangles, disks, or halfspaces.  
A seminal
paper by Bronniman and Goodrich~\cite{BronnGood}
introduced a multiplicative-weights method for
approximating \problem{Hitting Set}, and applied this
method to design constant-factor approximation algorithms
for various classes of bounded-VC-dimensional set systems,
e.g. disks in the plane.
The weighted case of these problems turns out to be much more
challenging; see~\cite{HarPeledLee,Varad10}.
A breakthrough paper by Mustafa and Ray~\cite{MustafaRay} 
presented a new method to analyze local search
algorithms for geometric hitting set problems,
thereby proving that local search yields a PTAS
for many interesting special cases such as
three-dimensional halfspaces.

The study of fixed-parameter approximation schemes --- 
and fixed-parameter
approximation algorithms more generally --- is still
in its youth.  An excellent survey by Marx~\cite{MarxApprox}
includes an fpt-AS for \mvcp (also known as \pvcp),
a problem which is a special case of bounded-VC-dimensional
\mcp.  Thus, one consequence of our algorithm in
Section~\svref{sec:fpt-as} is an alternative 
fpt-AS for \pvcp, although the techniques underlying
our algorithm are very different from those in
Marx's algorithm.

\section{Preliminaries}
\label{sec:prelim}

An instance of the \mcp problem is specified by a
finite set $\univ$, a collection of subsets $\rngspc$,
and a positive integer $k$.  We will assume that the
input is specified by simply listing the elements of
$\univ$ and those of each set in $\rngspc$; thus, the
problem size is $n = \sum_{R \in \rngspc} |R|$.
In the \problem{Weighted} \mcp problem, we are also given
a function $w : \univ \to \R_+$; the weight of a 
set $S \subseteq \univ$ is defined to be 
$w(S) = \sum_{x \in S} w(x)$ and the goal is
to output a $k$-tuple of elements of $\rngspc$
whose union has maximum weight.  We will denote
this maximum by $\opt(\univ)$.

For $A \subseteq \univ$, we will use the notation
$\rngspc|_A$ to denote the collection of all subsets
$B \subseteq A$ of the form $B = A \cap R$, where
$R \in \rngspc$.  The set $A$ is \emph{shattered}
by $\rngspc$ if $\rngspc|_A$ is equal to $2^A$,
the collection of all subsets of $A$.  The
VC-dimension of $(\univ,\rngspc)$ is the cardinality
of the largest set that is shattered by $\rngspc$.
If $(\univ,\rngspc)$ has VC-dimension $d$ and
$A \subseteq \univ$, it is known that 
$|\rngspc|_A|$ is bounded above by $O \left( |A|^d \right)$.

Our focus will be on \mcp problems such that
$(\univ,\rngspc)$ has bounded VC-dimension.
Among these, two special cases of particular
interest are \mhcp --- in which $\univ$ is a 
subset of $\R^d$ and each of the sets 
in $\rngspc$ is the intersection of a 
halfspace with $\univ$ --- and \mrcp,
in which $\univ$ is again a subset of $\R^d$
and each of the sets in $\rngspc$ is obtained by
intersecting an axis-parallel rectangle with $\univ$.

\section{A fixed-parameter approximation scheme}
\svlabel{sec:fpt-as}

In this section, we work with the \emph{unweighted}
\mcp problem.
Following Chazelle and Matou\v{s}ek, we assume
that $\rngspc$ is represented by
a \emph{subsystem oracle of dimension $d$},
defined as follows.

\begin{definition}[\cite{ChazMat}] \svlabel{def:oracle}
A \emph{subsystem oracle of dimension $d$} for a set
system $(\univ,\rngspc)$ is an algorithm which, given 
a subset $A \subseteq \univ$, returns a list of all
sets in $\rngspc|_A$ 
in time $O \left( |A|^{d+1} \right)$;
the number of sets in this list must also be bounded
above by $O \left( |A|^d \right)$.
\end{definition}

The following fact is obvious but useful: a subsystem oracle
of dimension $d$ for $(\univ,\rngspc)$ also constitutes
a subsystem oracle of dimension $d$ for $(V,\rngspc|_V)$,
for every subset $V \subseteq U$.

We define a set $A \subseteq \univ$ to be an 
$\eps$-approximation of $(\univ,\rngspc)$ if 
the inequality
\[
\left|
\frac{|A \cap R|}{|A|} - \frac{|R|}{|\univ|}
\right| \leq \eps
\]
holds for all $R \in \rngspc$.  A crucial ingredient
of our approximation scheme is an
algorithm, due to Chazelle and Matou\v{s}ek~\cite{ChazMat},
that computes an $\eps$-approximation of cardinality
$O \left( d \eps^{-2} \log(d/\eps) \right)$
for $(\univ,\rngspc)$ in time $O \left( d^{3d} \eps^{-2d} \log^d(d/\eps) n
\right)$, given a subsystem oracle of dimension $d$ for
$(\univ,\rngspc)$.

Let $\ku{\rngspc}$ denote the collection of all sets
$R_1 \cup \cdots \cup R_k$ such that $R_1,\ldots,R_k \in \rngspc$.
To apply Chazelle and Matousek's algorithm, we will need
a subsystem oracle for $\ku{\rngspc}$.  The existence of
such an oracle is ensured by the following lemma.

\begin{lemma} \svlabel{lem:oracle}
If $(\univ,\rngspc)$ has a subsystem oracle of dimension $d$,
then for all $k>0$, 
$(\univ,\ku{\rngspc})$ has a subsystem oracle of dimension
$kd$.
\end{lemma}
\svapdx{
The proof is given in Appendix~\ref{sec:fpt-as}.
}
{
\begin{proof}
The proof is by induction on $k$, the base case $k=1$ being trivial.
Given subsystem oracles for $(\univ,\rngspc)$ and
$(\univ,\kuv{\rngspc}{k-1})$ of dimensions $d$ and
$(k-1)d$, respectively, the following simple algorithm
constitutes a subsystem oracle for
$(\univ,\ku{\rngspc})$.  First, we use the given
two subsystem oracles to generate
lists $\mathcal{Q}_1$ and $\mathcal{Q}_{k-1}$,
consisting of all sets in $\rngspc|_A$ and
$\kuv{\rngspc}{k-1}|_A$, respectively.
Letting $a = |A|$, the
induction hypothesis implies that 
$|\mathcal{Q}_1| = O(a^d)$ and 
$|\mathcal{Q}_{k-1}| = O(a^{(k-1)d})$,
and that the two lists are generated in
time $O(a^{d+1})$ and $O(a^{(k-1)d+1})$,
respectively.  Now, for every pair $B_1 \in \mathcal{Q}_1$
and $B_{k-1} \in \mathcal{Q}_{k-1}$, we form the 
set $B = B_1 \cup B_{k-1}$ and add it to 
$\mathcal{Q}_k$.  There are 
$O(a^{kd})$ such pairs, and for each
pair the union can be computed in $O(a)$
time, so the algorithm runs in time $O(a^{kd+1})$,
as desired.\BKnote{If we need to form a list of sets
without duplicates, can that still be done in 
$O(a^{kd+1})$?  Probably, but it's not obvious 
to me.}
\end{proof}
}

As an easy consequence, we derive that the
\mcp problem has a fpt-AS
when $(\univ,\rngspc)$ has a bounded-dimensional
subsystem oracle \emph{and} the optimum is a 
constant fraction of $|\univ|$.

\begin{lemma} \svlabel{lem:easy-case}
For any constants $c,\delta>0$, consider 
the \mcp problem, restricted to set
systems $(\univ,\rngspc)$
having a subsystem oracle of dimension $d$ 
and satisfying $\opt(\univ) \geq c |\univ|$.
This special case of the \mcp
problem has a $\left( \frac{c-2\delta}{c} \right)$-approximation algorithm
with running time bounded by
$O(d^{3kd} k^{3kd} \delta^{-2kd-2} \log^{kd+1}(kd/\delta) n)$.
\end{lemma}
\begin{proof}
The set system $(\univ,\ku{\rngspc})$ has a 
subsystem oracle of dimension $kd$, so it is
possible to compute a set $A \subseteq \univ$
which is a $\delta$-approximation to $(\univ,\ku{\rngspc})$,
in time $O(d^{3kd} k^{3kd} \delta^{-2kd} \log^{kd}(kd/\delta) n)$.
Furthermore, the cardinality of $A$ is 
$O(k d \delta^{-2} \log(kd/\delta))$.
We can solve the \mcp problem
for the set system $(A,\rngspc|_A)$ by brute force.
First we call the subsystem oracle to obtain a list
of all the sets in $\rngspc|_A$; there are at most
$O(k^{d} d^d \delta^{-2d} \log^d(kd/\delta))$ such
sets.  Then we enumerate all $k$-tuples of sets
in this list, compute their union, and output 
the $k$-tuple whose union has the largest cardinality.
Computing the union of $k$ sets requires 
$O(k|A|) = O(k^2 d \delta^{-2} \log(kd/\delta))$ time,
and multiplying this by the number of $k$-tuples
we obtain an overall running time of
$O(k^{kd+2} d^{kd+1} \delta^{-2kd-2} \log^{kd+1}(kd/\delta))$.

Let $R_1,\ldots,R_k$ be sets in $\rngspc$ whose restrictions
to $A$ constitute an optimal solution of the 
\mcp problem for $(A,\rngspc|_A)$.
Let $S_1,\ldots,S_k$ be an optimal solution of
the \mcp problem for $(\univ,\rngspc)$.
We have
\begin{align*}
\frac{|(R_1 \cup \cdots \cup R_k) \cap A|}{|A|} & \geq
\frac{|(S_1 \cup \cdots \cup S_k) \cap A|}{|A|} \\
\frac{|R_1 \cup \cdots \cup R_k|}{|\univ|} & \geq
\frac{|S_1 \cup \cdots \cup S_k|}{|\univ|} - 2 \delta \\
\frac{|R_1 \cup \cdots \cup R_k|}{|S_1 \cup \cdots \cup S_k|} & \geq
1 - 2 \delta \left( \frac{|\univ|}{|S_1 \cup \cdots \cup S_k|} \right) \geq
1 - \frac{2 \delta}{c} = \frac{c - 2 \delta}{c}
\end{align*}
where the first line follows from the construction of
$R_1,\ldots,R_k$, the second line follows from the fact
that $A$ is an $\delta$-approximation for
for $(\univ,\ku{\rngspc})$, and the third line follows
from our assumption that $|S_1 \cup \cdots \cup S_k| = \opt(\univ) 
\geq c |\univ|$.
\end{proof}

For the remainder of this section, we work on eliminating
the assumption that $\opt(\univ) \geq c |\univ|$.  Our plan of 
attack is to perform a preprocessing step that extracts a
subset $V \subseteq \univ$ such that 
$\opt(V) \geq (1-\eps/3)\opt(\univ)$ and 
$\opt(V) \geq c |V|$, for a constant $c = c(\eps,k)$
depending only on $\eps$ and $k$.  Then we will run 
the algorithm from Lemma~\svref{lem:easy-case}
on $(V,\rngspc|_V)$, using an appropriate
choice of $\delta = \delta(\eps,k)$, 
to obtain a $(1-\eps)$-approximation
to $\opt(\univ)$.

To design and analyze the preprocessing algorithm that
constructs $V$, we must first define a new problem
that we call \cmcp
and analyze a simple greedy algorithm for the problem.

\begin{definition} \svlabel{def:constr-max-k-cov}
An instance of the \cmcp
problem is specified by a universe $\univ$ and 
$k$ collections of sets $\rngspc_1,\ldots,\rngspc_k \subseteq 2^{\univ}$.
A solution of the problem is specified by designating
a $k$-tuple of sets $R_1,\ldots,R_k$ such that $R_i \in \rngspc_i$ 
for $i=1,\ldots,k$.  The objective  is to maximize
$|R_1 \cup \cdots \cup R_k|$.

The greedy algorithm for \cmcp
selects $R_1,R_2,\ldots,R_k$, in that order, by choosing
$R_1$ to be the maximum-cardinality set in $\rngspc_1$ and,
for $i > 2$, choosing $R_i$ to be the set in $\rngspc_i$
that maximizes $|R_i \setminus (R_1 \cup \cdots \cup R_{i-1})|$.
\end{definition}

Note that \mcp is the special case
of \problem{constrained maximum $k$-coverage} in which 
$\rngspc_1 = \cdots = \rngspc_k$, and that the greedy
algorithm specializes, in that case, to the familiar greedy
algorithm for maximum $k$-coverage.  The approximation 
ratio of the greedy algorithm for \problem{constrained
maximum $k$-coverage} is not equal to $1-\frac{1}{e}$ in
general; in fact it is equal to $\frac12$.  However, for our 
purposes the following property of the greedy algorithm
will be more crucial to the analysis.

\begin{lemma} \svlabel{lem:greedy}
Given an instance of the \cmcp 
problem,
let $R_1,\ldots,R_k$ be the sets selected by the 
greedy algorithm and let $S_1,\ldots,S_k$ be any
other solution.  Let $\mathbf{R} = R_1 \cup \cdots \cup R_k$
and $\mathbf{S} = S_1 \cup \cdots \cup S_k$.
For every $\delta>0$, at least one
of the following two alternatives holds.
\begin{enumerate}
\item \svlabel{greedy:1}
$|\mathbf{R}| \geq (1-\delta) |\mathbf{S}|$.
\item \svlabel{greedy:2}
$|\mathbf{S} \setminus \mathbf{R}| < (1-\delta) |\mathbf{S}|$.
\end{enumerate}
\end{lemma}
\begin{proof}
We will construct a one-to-one mapping from
$\mathbf{S} \setminus \mathbf{R}$ into $\mathbf{R}$.
This suffices to prove the lemma, since either
$|\mathbf{S} \setminus \mathbf{R}| < (1-\delta) |\mathbf{S}|$
or $|\mathbf{S} \setminus \mathbf{R}| \geq (1-\delta) |\mathbf{S}|$,
and in the latter case our one-to-one mapping
will certify that 
$$ 
|\mathbf{R}| \geq |\mathbf{S} \setminus \mathbf{R}| \geq 
(1-\delta) |\mathbf{S}|.
$$

To construct the one-to-one mapping, partition
$\mathbf{S} \setminus \mathbf{R}$ into $k$ sets
$T_1, T_2, \ldots, T_k$, where 
$T_i = S_i \setminus (\mathbf{R} \cup S_1 \cup S_2 \cup \cdots \cup S_{i-1})$.
Note that $T_i$ is a subset of $S_i \setminus (R_1 \cup \cdots \cup R_{i-1})$,
hence 
$$
|T_i| \leq |S_i \setminus (R_1 \cup \cdots \cup R_{i-1})|
\leq |R_i \setminus (R_1 \cup \cdots \cup R_{i-1})|,
$$
where the second inequality follows from the definition of the greedy
algorithm.
This means that there is a one-to-one mapping
from $T_i$ to $R_i \setminus (R_1 \cup \cdots \cup R_{i-1})$.
Combining these one-to-one mappings gives us the
desired one-to-one mapping from
$\mathbf{S} \setminus \mathbf{R} = \coprod_{i=1}^k T_i$ 
into 
$\mathbf{R} = \coprod_{i=1}^k R_i \setminus (R_1 \cup \cdots \cup R_{i-1})$.
\end{proof}

We now describe and analyze a non-deterministic algorithm
to solve \mcp on a set system 
$(\univ,\rngspc)$, given a subsystem oracle of dimension $d$;
later we will make the algorithm deterministic.
The algorithm proceeds in a sequence of phases 
numbered $1,\ldots,p = \left\lceil 
\frac{6}{\eps} \ln \left( \frac{3}{\eps} \right) \right\rceil$.
In each phase $q$, it chooses a $k$-tuple of sets 
$R^q_1, \ldots, R^q_k$.  Let
$$
\mathbf{R}^q = \bigcup_{i=1}^q \bigcup_{j=1}^k R^i_j.
$$
In phase $q$, the algorithm computes a set $A^q$, of 
cardinality $O( k d \eps^{-2} p^2 \log(kdp/\eps) )$, which
is an $(\eps/6p)$-approximation to $(\mathbf{R}^{q-1},\ku{\rngspc})$.
It non-deterministically guesses a sequence of $k$
subsets $B^q_1,\ldots,B^q_k \subseteq A^q$ and defines 
set systems $\rngspc^q_1,\ldots,\rngspc^q_k$ as
\[
\rngspc^q_i = \{ R \in \rngspc \mid R \cap A^q = B^q_i \}, \quad
i=1,\ldots,k.
\]
It then selects the sets $R^q_1,\ldots,R^q_k$ using
the greedy algorithm for 
\cmcp,
applied to the universe $\univ \setminus \mathbf{R}^{q-1}$
with set systems $\rngspc^q_1,\ldots,\rngspc^q_k$.
After repeating this process for 
$p = \left\lceil 
\frac{6}{\eps} \ln \left( \frac{3}{\eps} \right) \right\rceil$
phases, it defines $V = \mathbf{R}^p = \bigcup_{i=1}^p \bigcup_{j=1}^k R^i_j$.
Setting $c=1/p$ and
$$
\delta = \frac{\eps}{6p} \left( 1 + \frac{\eps}{3} \right)^{-1},
$$
so that $(c-2 \delta)/c \geq 1 - \eps/3$, it runs the algorithm of 
Lemma~\svref{lem:easy-case} on the set system $(V,\rngspc)$
to find a $(1-\eps/3)$-approximation to the optimum of the
\mcp problem for $(V,\rngspc)$.

We aim to prove that there exists an execution of this 
non-deterministic algorithm that yields a $(1-\eps)$-approximation
to the optimum of the \mcp problem
for $(\univ,\rngspc)$.    
If our algorithm produces a set $V$ satisfying
$\opt(V) \geq c |V| = |V|/p$ and 
$\opt(V) \geq (1-\eps/3) \opt(U)$,
then Lemma~\svref{lem:easy-case} ensures that 
we finish up by producing a $(1-\eps/3)$-approximation 
to $\opt(V)$, which will also
be a $(1-\eps/3)^2\geq (1-\epsilon)$-approximation to
$\opt(U)$.  
Proving that $\opt(V) \geq |V|/p$
is easy: $V = \mathbf{R}^p$ is the union of $p$ sets
$\mathbf{R}^q \setminus \mathbf{R}^{q-1}$, each of 
which has cardinality at most $\opt(V)$ since it 
can be covered by the $k$ sets $R^q_1,\ldots,R^q_k$.

To prove that there exists an execution yielding a set
$V$ such that $\opt(V)\geq (1-\eps/3)\opt(U)$, we 
use Lemma~\svref{lem:greedy}.  Let $S_1,\ldots,S_k$ 
denote an optimal solution of the \mcp
problem for $(\univ,\rngspc)$.  Consider the execution 
in which the algorithm's choice of $B^q_i$ is equal to
$S_i \cap A^q$ for every $q,i$.  There are two cases to
consider.
First, suppose that exists a phase 
$q$ such that
\begin{equation} \svlabel{eq:fpt-1}
|(R^q_1 \cup \cdots \cup R^q_k) \setminus \mathbf{R}^{q-1}| \geq 
\left( 1 - \frac{\eps}{6} \right)
|(S_1 \cup \cdots \cup S_k) \setminus \mathbf{R}^{q-1}|.
\end{equation}
Recall that $\rngspc^q_i = \{
R \in \rngspc \mid R \cap A^q = B^q_i \},$
and that we are assuming $B^q_i = A^q \cap S_i$.
Hence, we have $R^q_i \cap A^q = S_i \cap A^q$ 
for all $i$ and, consequently, 
$(R^q_1 \cup \cdots \cup R^q_k) \cap A^q = 
(S_1 \cup \cdots \cup S_k) \cap A^q$.
Using the fact that $A^q$ is an $(\eps/6p)$-approximation
for $(\mathbf{R}^{q-1},\ku{\rngspc})$, we now obtain
\begin{equation} \svlabel{eq:fpt-2}
|(R^q_1 \cup \cdots \cup R^q_k) \cap \mathbf{R}^{q-1}| \geq 
|(S_1 \cup \cdots \cup S_k) \cap \mathbf{R}^{q-1}| - 
\frac{\eps}{6p} |\mathbf{R}^{q-1}|.
\end{equation}
Letting $\mathbf{S}$ denote $S_1 \cup \cdots \cup S_k$,
we sum~\sveqref{eq:fpt-1} and~\sveqref{eq:fpt-2} to obtain
\begin{align} \nonumber
|R^q_1 \cup \cdots \cup R^q_k| & \geq
\left( 1 - \frac{\eps}{6} \right) |\mathbf{S} \setminus \mathbf{R}^{q-1}|
+ |\mathbf{S} \cap \mathbf{R}^{q-1}| - \frac{\eps}{6p} |\mathbf{R}^{q-1}|
\\ & =
|\mathbf{S}| - \frac{\eps}{6} |\mathbf{S} \setminus \mathbf{R}^{q-1}|
- \frac{\eps}{6p} |\mathbf{R}^{q-1}|.
\svlabel{eq:fpt-3}
\end{align}
Now, as above, $\mathbf{R}^{q-1}$ can be partitioned into
sets $\mathbf{R}^{i} \setminus \mathbf{R}^{i-1}, \, (i=1,\ldots,q-1)$,
each having cardinality at most $\opt(U) = |\mathbf{S}|$.
The number of pieces of the partition is $q-1 < p$, so 
$\frac{1}{p}|\mathbf{R}^{q-1}| \leq |S|$.  Substituting
this back into~\sveqref{eq:fpt-3}, we obtain
\begin{equation} \svlabel{eq:fpt-5}
\opt(V) \geq |R^q_1 \cup \cdots \cup R^q_k| \geq 
\left(1 - \frac{\eps}{6} - \frac{\eps}{6} \right) |S| =
\left(1 - \frac{\eps}{3} \right) \opt(U),
\end{equation}
as desired.

Finally, there remains the case that~\sveqref{eq:fpt-1} 
is not satisfied by any $q$.  Then
Lemma~\svref{lem:greedy} implies that 
\begin{equation} \svlabel{eq:fpt-6}
|(S_1 \cup \cdots \cup S_k) \setminus \mathbf{R}^q| <
\left(1 - \frac{\eps}{6} \right) 
|(S_1 \cup \cdots \cup S_k) \setminus \mathbf{R}^{q-1}|
\end{equation}
for all $q$.  Combining~\sveqref{eq:fpt-6} for $q=1,\ldots,p$,
we get that
\[
|(S_1 \cup \cdots \cup S_k) \setminus \mathbf{R}^p| < 
\left(1 - \frac{\eps}{6} \right)^p |S_1 \cup \cdots \cup S_k|
\leq
\frac{\eps}{3} |S_1 \cup \cdots \cup S_k|,
\]
which implies that
\[
|(S_1 \cup \cdots \cup S_k) \cap \mathbf{R}^p| > 
\left( 1 - \frac{\eps}{3} \right) |S_1 \cup \cdots \cup S_k|,
\]
and hence $\opt(V) \geq (1 - \eps/3) \opt(U)$
since $V = \mathbf{R}^p$.

To turn the non-deterministic algorithm into a deterministic one,
we simply run every possible execution of the non-deterministic
algorithm and output the best answer.  An execution of the 
non-deterministic algorithm is determined by the choice of
sets $B^q_i, \; (1 \leq q \leq p, \, 1 \leq i \leq k)$.
Recall that $B^q_i$ must be a subset of $A^q$ and that
$|A^q| = O( k d \eps^{-2} p^2 \log(kdp/\eps) )$.  Hence
if $N(k,d,\eps)$ denotes the number of executions of the non-deterministic
algorithm, it satisfies
\svapdx{
\begin{align*}
N(k,d,\eps) & =  \prod_{q=1}^p \prod_{i=1}^k 2^{|A_q|} \\
\log N(k,d,\eps)  \leq pk \cdot O( k d \eps^{-2} p^2 \log(kdp/\eps) ) 
& = O( k^2 d \eps^{-2} p^3 \log(kdp/\eps) ) 
 = O( k^2 d \eps^{-5} \log^2(kd/\eps) )
\end{align*}
}{
\begin{align*}
N(k,d,\eps) & =  \prod_{q=1}^p \prod_{i=1}^k 2^{|A_q|} \\
\log N(k,d,\eps) & \leq pk \cdot O( k d \eps^{-2} p^2 \log(kdp/\eps) ) \\
& = O( k^2 d \eps^{-2} p^3 \log(kdp/\eps) ) \\
& = \tilde{O}( k^2 d \eps^{-5} )
\end{align*}
}
\BKnote{In this spot, I should write the running-time analysis for a single
iteration, but I'll get around to it later.}
Each iteration runs in time $O( g(k,d,\eps) \cdot n)$
where $\log g(k,d,\eps) = O(kd \log(kd/\eps))$.
Hence, the algorithm's overall running time is 
$O( f(k,d,\eps) \cdot n)$ where $\log f(k,d,\eps) =
\log N(k,d,\eps) + \log g(k,d,\eps) = \tilde{O}(k^2 d \eps^{-5} )$.

In deriving this bound on the algorithm's running time, 
we have assumed that
$(\univ,\rngspc)$ has a subsystem oracle of dimension $d$.
If we instead assume that $(\univ,\rngspc)$ has VC dimension $d$
and is represented in the input by simply listing all the
elements of $\rngspc$, the running time increases by a 
factor of $n$.  This is because the trivial implementation
of a subsystem oracle --- computing $\rngspc_A$
by enumerating each set of $\rngspc$ and intersecting
it with $A$ --- has running time $O(|A|^{d+1} n)$, 
$n$ times slower than the bound required by the definition
of a subsystem oracle.  


\section{Bounded Covering Multiplicity} \svlabel{sec:boundedCardinality}
In this section we show that the greedy algorithm gives a $1-(1-1/r)^r$-approximate solution when the \emph{covering multiplicty} of the set system is at most $r$. 
\begin{definition}
An instance of the maximum coverage problem 
$(\univ,\rngspc,k)$ has \emph{covering multiplicity} $r$
if for every $k$-tuple of sets $a_1,\ldots,a_k \in \rngspc$
there exists an optimal solution $(o_1,\ldots,o_k)$ of the
maximum coverage problem, with union denoted by $\opt$, 
such that each of the sets $a_i\cap \opt$ for
$1 \leq i \leq k$ is contained in the union of $r$ elements
of $\{o_1,\ldots,o_k\}$.
\end{definition}
One of the interesting special cases which satisfies this property
is when the cardinality of every set in $\rngspc$ is bounded by $r$.
In Appendix~\ref{sec:2d} we prove that it is also satisfied
(with $r=2$) when $\univ \subset \R^2$ and $\rngspc$ 
consists of halfspaces in $\R^2$.

Let $g_1,g_2,\ldots,g_k$ be the $k$ sets choosen by the  greedy algorithm in the order that they are choosen. Let $w$ be the coverage function and $o_1,o_2,\ldots,o_k$ be the $k$ sets choosen by OPT.
\begin{theorem}\svlabel{thm:boundedmultiplicity}
Greedy algorithm is a $1-(1-1/r)^r$ approximation algorithm for \mcp with \emph{covering multiplicity} $r$.
\end{theorem}
\begin{corollary}\svlabel{cor:boundedcardinality}
Greedy algorithm is a $1-(1-1/r)^r$ approximation algorithm for \mcp with each set having cardinality at most $r$.
\end{corollary}

\subsection{Reduction to a special case}
For simplying the analysis we first argue that we can consider the following special case without loss of generality. We take the problem instance on which the greedy algorithm (which we henceforth abbreviate as ``greedy'') has a given approximation ratio and convert it into a special instance with no better approximation ratio. Then it is enough to analyze the special case.
\begin{itemize}
\item The sets chosen by greedy are different from the optimal sets. This assumption can be made as we can just duplicate the sets. Note that this does not change the {covering multiplicity}.
\item The sets chosen by greedy are disjoint. This is because if one defines a new problem instance with $\tilde{g}_i=g_i \setminus \left(\cup_{j=1}^{i-1}g_{j}\right)$ then the values of the optimal solution and the greedy solution are unchanged.  Note that this step uses the fact that the sets chosen by greedy do not belong to the optimal solution.  Also note that this does not change the {covering multiplicity} since we are not modifying any sets in the optimal solution.
\item Let $o_1,o_2,\ldots,o_k$ be any optimal solution such that each set $g_i \cap OPT \; (1 \leq i \leq k)$ is contained in the union of $r$ elements
of $\{o_1,\ldots,o_k\}$.  We can assume that these sets $o_i$ are pairwise disjoint. This is because we can define a new problem instance in which every point belonging to two or more of the sets in $\{o_1,\ldots,o_k\}$ is assigned to one of those sets and deleted from the others.  The values of the greedy and optimal solutions are unchanged.  To preserve the property that each set $g_i \cap OPT \; (1 \leq i \leq k)$ is contained in the union of $r$ elements of $\{o_1,\ldots,o_k\}$, we simply ensure that every element of $g_i$ is assigned to one of those $r$ sets, for all $i$.  This is possible due to our previous assumption that the sets $g_1,\ldots,g_k$ are disjoint.
\end{itemize}

\subsection{Simple case}
Consider the simple case $k=t\cdot r$ for some integer $t$. We will prove the approximation for this special case to get some intuition. We will do it in steps.
\begin{itemize}
\item Let $x_i=\sum_{j=(i-1)\cdot t+1}^{i\cdot t}w(g_i)$.
\item Let $o_1,o_2,\ldots,o_k$ be the optimal sets 
in decreasing order of $w(o_i)$.
\item Note that $w(g_1)\geq w(o_1), \, w(g_2)\geq w(o_{r+1}), \, w(g_3)\geq w(o_{2r+1}),\ldots,\, w(g_t)\geq w(o_{(t-1)\cdot r+1})$. These inequalities use the fact that 
the covering multiplicity is $r$ and the sets $o_1,\ldots,o_k$ are disjoint.
Now summing the $t$ terms we get $\sum_{i=1}^t w(g_i)\geq \sum_{i=1}^t w(o_{(i-1)r+1})\geq \frac{1}{r}w(OPT)$.
\item Repeating the above step on the residual problem we get $\sum_{i=t+1}^{2t}w(g_i)\geq \frac{OPT-x_1}{r}$. Similarly we get the following series of equations.
\begin{eqnarray} \label{eq:simpleone}
 \forall 1\leq l\leq r-1, \sum_{i=l\cdot t+1}^{(l+1)t}w(g_i)\geq \frac{OPT-\sum_{i=1}^{l}x_i}{r}
\end{eqnarray}
\item Multiplying \eqref{eq:simpleone} by $(1-1/r)^{r-l-1}$ and summing we get $\sum_{i=1}^{i=k}w(g_i)\geq \left( 1- \left( 1- \frac1r \right)^r \right)w(OPT)$.
\end{itemize}

\subsection{General case}
Let $k=t\cdot r+q$ for some $0\leq q\leq r-1$. We will use the following lemma in the proof.
\begin{lemma} \label{eq:generallone}
$\forall 1\leq l\leq r,0\leq z\leq q\leq r$ we have $r\cdot \sum_{m=0}^z \binom{l-1}{m}\binom{r-l}{q-m-1}\geq q\cdot \sum_{m=0}^{z}\binom{l-1}{m}\binom{r-l+1}{q-m}$
\end{lemma}
\begin{proof}
Consider $p(z)=\frac{\sum_{m=0}^z \binom{l-1}{m}\binom{r-l}{q-m-1}}{\sum_{m=0}^{z}\binom{l-1}{m}\binom{r-l+1}{q-m}}$. Consider a random process in which $q$ out of $r$ bins are chosen uniformly at random (without replacement) and a ball is added to each one of the $q$ bins. Now $p(z)$ represents the conditional probability that a ball is in the $l^{th}$ bin, given that at most $z$ bins from the first $l-1$ are chosen.  One can easily see that this function should be a decreasing function of $z$ and hence $p(z)\geq p(q)=\frac{q}{r}$.
\end{proof}

Consider $r$ bins and arrange the $k$ greedy sets in the $r$ bins with each bin having either $t$ or $t+1$ greedy sets. Let bin 1 have the first $t$ or $t+1$ sets, bin 2 have the second $t$ or $t+1$ sets, and so on. Let $\sigma$ be one such arrangement. We will apply inequalities similar to the simpler case. Let $x_i^\sigma=\sum_{g_j\in bin_i} w(g_j)$. Let $\sigma(l)$ denote the number of sets in the first $l$ bins. Let $x_{qmin}^{t}$ be the residual value of the $q^{th}$ minimum set among the optimal sets after the first $t$ greedy sets are choosen. Note that $x_{qmin}^t$ is a decreasing function of $t$. Let $B(t)$ denote the set of bins with $t$ sets and $B(t+1)$ denote the set of bins with $t+1$ sets.
\begin{itemize}
\item Consider bin $l$ with $t+1$ items. Then $\sum_{g_i\in bin_l}w(g_i)\geq \frac{OPT-\sum_{i=1}^{l-1}x_i^\sigma+(r-q)x_{qmin}^{\sigma(l-1)}}{r}$. This inequality is proved similar to inequality $\ref{eq:simpleone}$.
\item Consider bin $l$ with $t$ items. Then $\sum_{g_i\in bin_l}w(g_i)\geq \frac{OPT-\sum_{i=1}^{l-1}x_i^\sigma-q\cdot x_{qmin}^{\sigma(l-1)}}{r}$. This inequality is proved similar to inequality $\ref{eq:simpleone}$.
\end{itemize}
Multiplying the equation corresponding to bin $l$ with $(1-1/r)^{r-l}$ and summing we get 
\begin{eqnarray}
w(greedy)&\geq& (1-(1-1/r)^r w(OPT) \nonumber \\
& & \;\; +\sum_{bin_l\in B(t+1)} \frac{r-q}{r}\cdot (1-1/r)^{r-l}x_{qmin}^{\sigma(l-1)}-\sum_{bin_l\in B(t)} \frac{q}{r}\cdot (1-1/r)^{r-l}x_{qmin}^{\sigma(l-1)}
\end{eqnarray}
Now taking the average over all arrangements $\sigma$ we get the following equation.
{\small
\begin{eqnarray}
w(greedy)&\geq& (1-(1-1/r)^r w(OPT) \nonumber \\
& & \;\; +\sum_l \frac{(1-1/r)^{r-l}}{\binom{r}{q}} \left(\frac{r-q}{r}\sum_{w=0}^{l-1} \binom{l-1}{w}\binom{r-l}{q-w-1}x_{qmin}^{(l-1)t+w}-\frac{q}{r}\sum_{w=0}^{l-1}\binom{l-1}{w}\binom{r-l}{q-w}x_{qmin}^{(l-1)t+w} \right) \nonumber \\
&\geq&(1-(1-1/r)^r w(OPT) \nonumber \\
& & \;\; +\sum_l \frac{(1-1/r)^{r-l}}{r\cdot \binom{r}{q}} \left(\sum_{w=0}^{l-1}(((r-q)\binom{l-1}{w}\binom{r-l}{q-w-1}-q\binom{l-1}{w}\binom{r-l}{q-w})x_{qmin}^{(l-1)t+w} \right) \nonumber \\
&\geq&(1-(1-1/r)^r w(OPT) \nonumber \\
& & \;\; +\sum_l \frac{(1-1/r)^{r-l}}{r\cdot \binom{r}{q}} \left(\sum_{w=0}^{l-1}((r\binom{l-1}{w}\binom{r-l}{q-w-1}-q\binom{l-1}{w}\binom{r-l+1}{q-w})x_{qmin}^{(l-1)t+w} \right)
\end{eqnarray}
}
Now using the fact that $x_{qmin}^t$ is a decreasing function of $t$ and Lemma \ref{eq:generallone} we get $w(greedy)\geq (1-(1-1/r)^r) w(OPT)$.

\section{Lower bounds}
\svlabel{sec:lb}

This section considers three different low-dimensional
restrictions of \mcp: set systems of VC-dimension 2,
halfspaces in $\R^4$, and axis-parallel rectangles in $\R^2$.
In each case, we show that the problem is APX-hard and that
the greedy algorithm's approximation ratio, restricted to
that special case, is no better than its worst-case
approximation ratio, $1 - \frac1e$.

All of these lower bounds are based on the \mvcp problem,
the special case of \mcp in which each element of $\univ$
belongs to exactly two sets in $\rngspc$.  In this special
case, we can identify $\rngspc$ with the vertex set of 
a graph $G$, and $\univ$ with its edge set, such that the
endpoints of the edge corresponding to $x \in \univ$ are
the vertices that correspond to the two sets containing $x$.
Thus, \mvcp can be defined as the problem of choosing
$k$ vertices of a graph to maximize the number of edges 
they cover.  The problem is known to be
APX-hard~\cite{Petrank} and it is known 
that the approximation ratio of the 
greedy algorithm, specialized to 
\mvcp, is no better than in the general case~\cite{CNW80}.
In fact, the following lemma shows that 
the performance of the greedy algorithm
does not improve when we further specialize to
bipartite instances of \mvcp.
\begin{lemma} \svlabel{lem:mvcp}
For any $\eps>0$, 
there exist instances
of \mvcp in which the graph is bipartite,
the instance has a vertex cover of size $k$,
but the output of the greedy algorithm 
covers only $1 - \frac1e + \eps$ fraction
of the edges.
\end{lemma}
The proof consists of taking a well-known hard
example for the greedy \mcp algorithm, and
encoding it in the form of a bipartite graph;
the details are given in Appendix~\ref{subsec:missingProofs}.

\begin{theorem} \svlabel{thm:lb}
Each of the following special cases of \mcp is APX-hard:
\begin{compactenum}[(a)]
\item \svlabel{lb:vc}  Set systems of VC-dimension $d \geq 2$.
\item \svlabel{lb:hlf} Halfspaces in $\R^d, \; d \geq 4$.
\item \svlabel{lb:rct} Rectangular ranges in $\R^d, \; d \geq 2$.
\end{compactenum}
Furthermore, the worst-case approximation ratio of the 
greedy algorithm, when restricted to any of these special
cases, is $1 - \frac1e$.
\end{theorem}
\begin{proof}[Proof Sketch]
The full details of the proof 
are given in Appendix~\ref{subsec:missingProofs}.
Part~(\ref{lb:vc}) is a restatment of the known results on \mvcp.
To show Part~(\ref{lb:hlf}) we embed \mvcp into halfspaces in $\R^d, \; d\geq 4$
to reconstruct similar results. For Part~(\ref{lb:rct}) to show the APX-hardness, 
we use a reduction from
\problem{Bounded-Degree Vertex Cover}, which
was shown to be APX-hard by Papadimitriou and
Yannakakis~\cite{PY91}.  For the statement about
the approximation ratio of the greedy algorithm,
we use Lemma~\svref{lem:mvcp}.
\end{proof}

An immediate corollary of Theorem~\svref{thm:lb} is the
following statement, which justifies that in our fixed-parameter
algorithm, the super-polynomial dependence of the running time on 
$k$ and $1/\eps$ is unavoidable.
\begin{corollary} \svlabel{cor:fpt-as}
Suppose that \mcp, specialized to instances with VC-dimension $d$, 
has a $(1-\eps)$-approximation algorithm with running
time $O(f(\eps,k,d) \cdot \poly(n))$, for every $\eps,k$.  
If $P \neq NP$, then $f(\eps,k,d)$ must be super-polynomial in $k$.
If $P \neq W[1]$, then $f(\eps,k,d)$ must be super-polynomial in $\eps^{-1}$.
In fact, both of these statements hold even if we restrict to $d=2$.
\end{corollary}
\begin{proof}
The statement that $f(\eps,k,2)$ must be super-polynomial in $k$
is a restatement of the APX-hardness of \mcp in VC-dimension 2,
which is Part~(\svref{lb:vc}) of Theorem~\svref{thm:lb}.
To prove that $f(\eps,k,2)$ must be super-polynomial in
$\eps^{-1}$, we observe that \mcp, specialized to instances with
VC-dimension 2, is a generalization of the $W[1]$-hard
\problem{partial vertex cover} problem,
and that approximating the optimum of \problem{partial vertex cover} 
within a factor of $(1-\eps)$, for $\eps < 1/|E|$, is equivalent to 
solving it exactly.
\end{proof}

\section{Open Questions}
We leave several interesting open questions.
\begin{itemize}
\item Improve the running time of our algorithm for sets with bounded VC-dimension.
\item Give an algorithm better than $1-(1-1/r)^r$ approximation when the cardinality of each
set is bounded by $r$. Such an algorithm could have a running time exponential in $r$.
\item Resolve the approximability of max-coverage on $3$-dimensional halfspaces. We conjecture 
that local search is a PTAS for the problem.  Appendix~\ref{sec:2d-ptas}
 presents a proof of the two-dimensional version of this conjecture.
\end{itemize}

\bibliographystyle{plain}
\bibliography{ldcp}

\appendix
\section{Appendix}

\subsection{Missing Proofs from Section~\ref{sec:lb}}\label{subsec:missingProofs}
\begin{lemma} 
For any $\eps>0$, 
there exist instances
of \mvcp in which the graph is bipartite,
the instance has a vertex cover of size $k$,
but the output of the greedy algorithm 
covers only $1 - \frac1e + \eps$ fraction
of the edges.
\end{lemma}
\begin{proof}
We construct a bipartite graph with edge set
$E = \{1,\ldots,kN\}$ (for some sufficiently large $N$)
and vertex set $U \cup W$, where $k = |U| \leq |W|$.
We refer to $U$ and $W$ as the \emph{left} and \emph{right}
vertex sets, respectively.

Define a sequence of positive integers $n_0, n_1, n_2, \ldots$ by the
formula $n_i = \left\lceil N \cdot \left( 1 - \frac1k \right)^{i}
\right\rceil + 1$ and let $s_i = \sum_{j=0}^{i-1} n_j$ denote the
sequence of partial sums, interpreting $s_0$ to be 0.  
If $r = \min \{ i \mid s_{i} \geq kN \}$ then 
$W = \{w_1, \ldots, w_r\}$, while $U = \{u_1,\ldots,u_k\}$.
The right endpoint
of edge $j$ is the unique $w_i$ such that $s_{i-1} < j \leq s_{i}$,
while the left of endpoint of $j$ is the unique $u_i$ such that
$i \equiv j \pmod{k}$.

By construction, $|U|=k$ and $U$ is a vertex cover.  Each element
of $U$ has exactly $N$ elements.  
However, the greedy algorithm instead chooses vertices $w_1,\ldots,w_k$.
To prove this by induction, observe that
after choosing $w_1,\ldots,w_i$, the
number of remaining uncovered edges is less than $kN \left(
1 - \frac1k \right)^{i}$, and these edges are consecutively
numbered.  Each element of $U$ covers a congruence class of 
edges, and therefore it covers fewer than 
$N \left(1 - \frac1k \right)^{i} + 1$ of the remaining edges,
whereas $w_{i+1}$ covers $n_i$ edges and 
$n_i \geq N \left(1 - \frac1k \right)^{i} + 1$.
It follows that the greedy algorithm chooses $w_{i+1}$ and
this completes the induction step.

The number of edges covered by $w_1,\ldots,w_k$
is bounded above by $2k + N \sum_{i=0}^{k-1} \left(1 - \frac1k \right)^i =
2k + kN \left[ 1 - \left( 1 - \frac{1}{k} \right)^{k} \right]$.
For $k,N$ sufficiently large, this is less than
$\left(1 - \frac1e + \eps\right) k N$.
\end{proof}

\begin{theorem} \svlabel{thm:app:lb}
Each of the following special cases of \mcp is APX-hard:
\begin{compactenum}[(a)]
\item \svlabel{lb:app:vc}  Set systems of VC-dimension $d \geq 2$.
\item \svlabel{lb:app:hlf} Halfspaces in $\R^d, \; d \geq 4$.
\item \svlabel{lb:app:rct} Rectangular ranges in $\R^d, \; d \geq 2$.
\end{compactenum}
Furthermore, the worst-case approximation ratio of the 
greedy algorithm, when restricted to any of these special
cases, is $1 - \frac1e$.
\end{theorem}
\begin{proof}
Recall that \mvcp can be defined as the instance of
\mcp in which every $x \in \univ$ belongs to exactly
two sets in $\rngspc$.  Any such set system $(\univ,\rngspc)$
has VC-dimension at most 2: indeed, if $\rngspc$ shatters 
a three-element set $\{x,y,z\}$ then there exist sets
$R_1,\ldots,R_4$ in $\rngspc$ whose intersections with
$\{x,y,z\}$ are the sets $\{x\}, \, \{x,y\}, \, \{x,z\}, \,
\{x,y,z\}$, respectively, and consequently $x$ belongs to 
at least four distinct sets in $\rngspc$.  Thus, we see
that \mcp restricted to set systems of VC-dimension $d$
includes \mvcp as a special case, as long as $d \geq 2$. 
Part~(\svref{lb:app:vc}) of
the theorem now follows from the fact that \mvcp is 
APX-hard~\cite{Petrank} and from Lemma~\svref{lem:mvcp}.

To prove Part~(\ref{lb:app:hlf}) we again show that \mvcp is
a special case.  To do so, consider any graph with vertex
set $V = \{v_1,\ldots,v_n\}$ and associate to each vertex 
$v_t \in V$ the vector $b_t = (t, t^2, t^3, t^4, 0, \ldots, 0) \in \R^d$.
Define a halfspace $\hlf_t \subset \R^d$ by the inequality 
$v_t \cdot x \geq 1$.  For
every edge $(v_r,v_s)$
we construct a vector $y_{rs} \in \R^d$ that belongs to 
$\hlf_r \cap \hlf_s$ but not to $\hlf_t$ for any 
$t \neq i,j$.  The construction is as follows.  First,
write the polynomial $(z-r)^2 (z-s)^2$ in the form
$\sum_{i=0}^4 a_i z^i$, and then put
$$y_{rs} = - \frac{1}{a_0} ( a_1, a_2, a_3, a_4, 0, \ldots, 0) \in R^d.$$
The inequality $y_{rs} \cdot v_t \geq 1$ can be rewritten as
$- \sum_{i=1}^4 a_i t^i \geq a_0$ (using the fact that $a_0 = r^2 s^2 > 0$)
and it follows that the inequality is satisfied only when
$(t-r)^2 (t-s)^2 \leq 0$, i.e.~only when $t \in \{r,s\}$.
Thus, the set system defined by the vectors $\{y_{rs}\}$
and the halfspaces $\{\hlf_t\}$ is identical to the \mvcp
instance defined by $G$.

To prove Part~(\ref{lb:app:rct}), we specialize to rectangular
ranges in $\R^2$.  (The case of rectangular 
ranges in $\R^d, \, d>2$ follows \emph{a fortiori}.)
To begin with, we observe that 
every \emph{bipartite} instance of \mvcp can be represented
using axis-parallel rectangles in $\R^2$.  
The construction is as follows.
If we label the vertices of the bipartite graph as $\{u_1, \ldots, u_p,
w_1, \ldots, w_q\}$ such that every edge has one endpoint 
in $\{u_1,\ldots,u_p\}$ and the other endpoint in $\{w_1,\ldots,w_q\}$,
then we can represent edge $(u_i,w_j)$ using the point $(2i,2j) \in \R^2$.
Vertex $u_i$ is represented by the rectangle
$[2i-1, 2i+1] \times [1,2q+1]$
and vertex $w_j$ by the rectangle
$[1,2p+1] \times [2j-1, 2j+1]$.
This construction, combined with Lemma~\ref{lem:mvcp},
suffice to show that the greedy algorithm has worst-case
approximation ratio $1 - \frac1e$ when specialized
to rectangular ranges in $\R^2$.  To prove APX-hardness,
we need to use a different reduction that is based on
bounded-degree graphs rather than bipartite graphs.
We use the following theorem from~\cite{PY91}:
there exists a constant $\Delta$ such that 
\problem{vertex cover}, restricted to graphs
of maximum degree $\Delta$, is APX-hard.  

For any graph $G$, create an instance of
\mcp as follows.  Assuming that $G$ has vertex
set $\{v_1,\ldots,v_n\}$ and edge
set $\{e_1,\ldots,e_m\}$.  
For each edge $e_k$ with endpoints
$v_i, v_j$, the set $\univ \subset \R^2$
contains the three points 
$(6k-2,2i), \, (6k,0), \, (6k+2, 2j)$.
These $3m$ points constitute the entire set $\univ$.
The rectangles in $\rngspc$ are as follows.
For each edge $e_k$ there are two rectangles
$\rct_1(e_k) = [6k-3,6k+1] \times [-1,n+1]$
and $\rct_2(e_k) = [6k-1,6k+3] \times [-1,n+1]$.
For each vertex $v_i$ there is one rectangle
$\rct(v_i) = [0,6m+3] \times [2i-1,2i+1]$.
If $G$ has a vertex cover $C$ of size $s$, then
there is a set of $m+s$ rectangles in $\rngspc$
that cover all the points in $\univ$: we take
rectangle $\rct(v_i)$ for each $v_i \in C$,
this covers at least one of the points 
$(6k-2,2i), \, (6k+2,2j)$ for each edge $e_k$
and the remaining two points corresponding to
that edge can be covered using either 
$\rct_1(e_k)$ or $\rct_2(e_k)$.  
Conversely, if $\univ$ can be covered
by $m+s$ elements of $\rngspc$, then
the covering must have a subcollection of $m$
rectangles that contains one
of the rectangles $\rct_1(e_k), \rct_2(e_k)$
for each $k$.  Let $T$ be the subset of
$\univ$ that is not covered by this subcollection,
and let $C$ be the set of all vertices $v_i$
such that $T$ contains a point whose $y$-coordinate
is $2i$.  It is easy to see that $C$ is a vertex
cover of $G$, and $|C| \leq s$.

Now let $\eps>0, \Delta < \infty$ be chosen such that 
it is NP-hard to distinguish between graphs of
maximum degree $\Delta$ having a vertex cover of
size $s$ (henceforth, \emph{yes instances}) 
and those having no vertex cover of size
less than $(1+\eps)s$ (\emph{no instances}).  
If $G$ is a yes instance, then the corresponding
\mcp instance with parameter $k=m+s$ has optimum
value $3m$.  If $G$ is a no instance, then the
corresponding $\mcp$ instance with parameter
$k=m+s$ has optimum value at most $3m - \eps s$.
Indeed, if there exist $m+s$ rectangles that cover
more than $3m - \eps s$ points, then it is trivial to find
fewer than 
$m+(1+\eps)s$ rectangles that cover all $3m$ points,
which is impossible if $G$ is a no instance.
If it possible for a graph with $m$ edges and maximum
degree $\Delta$ to have a vertex cover of size $s$
then $s \geq m/\Delta$.  Thus, we have shown that it 
a yes instance of \problem{vertex cover} maps to a
\mcp instance whose optimum value is $3m$ while a
no instance maps to one whose optimum value is 
at most $(3 - \eps/\Delta)m$, implying the claimed
APX-hardness.
\end{proof}

\section{Two-dimensional halfspaces}
\svlabel{sec:2d}

Theorem~\svref{thm:lb} rules out the possibility of 
designing a PTAS for \mcp specialized to halfspaces in $\R^d$ for
$d \geq 4$ (unless P=NP) and it likewise rules out the
possibility of proving an approximation ratio better than
$1 - \frac{1}{e}$ for the greedy algorithm.  
But in very low dimensions, the situation is different.
When $d=1$, it is easy to see that the greedy algorithm itself 
always computes an optimal solution.  
When $d=2$, a dynamic programming algorithm due to 
Har-Peled and Lee~\cite{HarPeledLee} computes an optimal
solution in polynomial time.  (The algorithm given in that
paper is for \problem{Set Cover} rather than \mcp, but 
a trivial modification of their algorithm solves \mcp.)
Despite the existence of a polynomial-time algorithm
for two-dimensional \mhcp, it is interesting to investigate
the approximation ratio of some other archetypical
algorithms for this problem, especially since this investigation 
may shed light on the approximability of three-dimensional
\mhcp, which is NP-hard~\cite{FG88} and hence the two-dimensional
dynamic programming algorithm is unlikely to generalize.
In this section, we show
that when $d=2$ the greedy algorithm has approximation ratio $3/4$,
and there is a natural local search algorithm yielding a PTAS.  



 \subsection{Analysis of the greedy algorithm}
\svlabel{sec:greedy}
To analyze the greedy algorithm for \mhcp in two dimensions we prove that
the \emph{covering multiplicity} $r$ of the problem instance is $2$. Then
by Theorem~\ref{thm:boundedmultiplicity} we have that greedy algorithm is a factor $\frac34$ approximation
algorithm for \mhcp in two dimensions.

\begin{lemma} \label{lemma:2dcovering}
The \emph{covering multiplicity} for \mhcp in two dimensions is $2$.
\end{lemma} 
\begin{proof}
The proof is a series of simple observations.  
\begin{enumerate}[(a)]
 \item Without loss of generality we can assume that no set belongs to both
 the optimal solution and the given solution $G$. 
 This is because otherwise we can duplicate the set and treat one copy as
 belonging to the optimal solution while the other belongs to the given
 solution.
 \item Consider the optimal solution which has the maximum 
 number of sets in common with the duplicates created in the previous step.
 \item In the optimal solution $O$ considered above, 
 for every other set $s \not\in O$, there are two sets $o_1,o_2 \in O$
such that every element of $s$ that is covered by $O$ belongs to $o_1 \cup o_2$.
Otherwise, using the fact that this is a two-dimensional \mhcp instance, we can see that one of the previous two conditions is violated.
 \end{enumerate}
\end{proof}
\begin{theorem} \svlabel{thm:2d-greedy}
The greedy algorithm for two-dimensional \mhcp has
approximation ratio 3/4.
\end{theorem}
\begin{proof}
The proof follows easily from Lemma~\ref{lemma:2dcovering} and Theorem~\ref{thm:boundedmultiplicity}.
\end{proof}

\begin{example} \svlabel{ex:greedy-tight}
The following example shows that the analysis of the greedy algorithm is tight.
Consider the set system $s_1=\{p_1,p_2\}$,$s_2=\{p_3,p_4\}$ and $s_3=\{p_1,p_3\}$ with $k=2$. Then it should be simple
to see that this can be realized an a two-dimensional
instance of \mhcp.
One choice for optimal sets is $s_1,s_2$ with value of $4$. One possible output for the greedy 
algorithm is $s_3,s_1$ with value $3$. This gives an approximation of $3/4$.
\end{example}

 \subsection{A PTAS via local search} \svlabel{sec:2d-ptas}

If $(\univ,\rngspc,k)$ is an instance of \mcp and 
$S = \{ R_1,\ldots,R_k \}$ is a solution, define a $t$-swap
to be the operation of transforming this solution into
another solution $S' = \{ R'_1, \ldots, R'_k \}$ such that 
there are at most $t$ sets belonging to $L$ but not $L'$,
and vice-versa.  
If $(\univ,\rngspc,k)$ is a two-dimensional instance of \mhcp
and $S = \{R_1,\ldots,R_k\}$ is a solution, define
$area(S) \subseteq \R^2$ to be the set $\bigcup_{i=1}^k \hlf_i$,
where $\hlf_i$ is the halfspace corresponding to $R_i$.

In this section we analyze the following local search algorithm.
We assume an \emph{unweighted} instance of two-dimensional
\mhcp, i.e. an instance in which each element has weight 1.
\svapdx{ \begin{compactenum} }{ \begin{enumerate} }
 \item Start with a arbitrary $k$-tuple of sets $S$.
 \item While possible do a $t$-swap to improve the number of elements
covered.
 \item If there exists a 1-swap to obtain a solution $S'$ such that $area(S)\subseteq area(S')$ and $area(S)\neq area(S')$ then perform this 1-swap and go to step 2.  Otherwise terminate the algorithm.
 \svapdx{ \end{compactenum} }{ \end{enumerate} }
 It is simple to see that step~3 does not run for more than $n$ times without the solution improving because once a set is deleted from $S$ in step~3, the only event that can re-insert it is a $t$-swap in step~2.

We will prove that this local search algorithm achieves an
approximation ratio of $2t/(2t+1)$.  (This implies that we
can obtain a PTAS with running time $n^{O(1/\eps)}$ by setting
$t = 1/\eps$.)  The proof of the approximation ratio is in two steps. 
We first assume the existence of a certain chain decomposition and 
prove that this implies a $2t/(2t+1)$ approximation ratio. 
Then we construct such a chain decomposition.
 
 \subsubsection{Approximation ratio assuming chain decomposition}
For succinctness, we will refer to the sets in the optimal solution 
and in the output of the local search algorithm as \emph{opt sets}
and \emph{local sets}, respectively.  Opt sets will be denoted by
$o_i$ and local sets by $\ell_i$.  If $L_i$ is a subcollection of
the local sets, we will frequently use the notation 
$v(L_i)$ to denote the set of elements covered by $L_i$ but not by
any of the other local sets, i.e.
$$
v(L_i) = \left( \bigcup_{\ell_j \in L_i} \ell_j \right) 
-
\left( \bigcup_{\ell_j \not\in L_i} \ell_j \right).
$$
Let the opt sets and local sets be grouped into chains $C_1,...,C_l$ such that the following properties are satisfied.
\begin{itemize}
\item the local and opt sets alternate (cyclically) in a chain $C_i$.
\item Consider a portion of any chain (cyclically) $...o_1\ell_1o_2...\ell_to_{t+1}...$. Let $L=\{\ell_1,\ell_2,...,\ell_t\}$, $O_1=\{o_1,o_2,...,o_t\}$ and $O_2=\{o_2,o_3,...o_{t+1}\}$ Then $v(L)\cap OPT\subseteq v(L)\cap (O_1\cup O_2)$.
\end{itemize}
\svapdx{
In Appendix~\ref{sec:2d} we prove the existence of such a chain
decomposition, and we show that it implies the claimed $2t/(2t+1)$
approximation.  The proof is elementary but the construction
of the chain decomposition is somewhat intricate.
}
{
Consider $L_i$ and $O_i$ each having same number of sets and at most $t$ sets. We derive some equations based on local optimality.
\begin{eqnarray} \svlabel{localone}
w(v(L_i)-OPT)+w(v(L_i)\cap OPT) &\geq& w(O_i-local)+w(O_i\cap v(L_i)) 
\\
\Rightarrow w(v(L_i)-OPT)+w(v(L_i)\cap (OPT-O_i)) &\geq& w(O_i-local)
\end{eqnarray}
Now we find sets used in equation \sveqref{localone} and then add these equations to get the desired result.
\begin{itemize}
\item Consider chain $C_i$. If number of local sets in $C_i$ is $\leq t$, then  let $L_i$ be the collection of all local sets in $C_i$ and let $O_i$ be the collection of all opt sets in $C_i$.  Make $2t$ such copies, i.e.~the same equation will be used $2t$ times in the proof.
\item If chain $C_i$ chain has more than $t$ local sets, then let $L_i$ be any $t$ consecutive local sets, and let $O_i$ be the opt sets in the chain which are shifted from $L_i$ by 1 either clockwise or counterclockwise.  Note that a particular choice for $L_i$ appears twice since there are two options for $O_i$.  
\end{itemize}
Here are some properties of the above decomposition.
\begin{enumerate}
\item Each $\ell_i$ belongs to $2t$ of the sets $L_j$
\item Each $o_i$ belongs to $2t$ of the sets $O_j$
\item Let $L_i$ be associated with $O_j$ and $O_k$. Then $v(L_i)\cap OPT\subseteq v(L_i)\cap (O_j\cup O_k)$. This is just a restatement of the assumed property of the chain decomposition.
\end{enumerate}
Based on the above properties we derive the the final inequality. Sum the equation \sveqref{localone} over all $L_i$,$O_i$. Then we bound each term in the sum.  Let $\opt$ denote the set of elements covered by the opt sets, and let $\loc$ denote the set of elements covered by the local sets.
\begin{itemize}
\item $\sum w(v(L_i)-\opt)\leq 2t\cdot w(\loc-\opt)$. This is due to property 1.
\item $\sum w(O_i-\loc)\geq 2t\cdot w(\opt-\loc)$. This is due to property 2. Note the difference in the direction of inequalities.
\item $\sum w(v(L_i)\cap (\opt-O_i))\leq w(\loc \cap \opt)$. This is due to property 3.
\end{itemize}
From the above three equations and equation \sveqref{localone} we get the final necessary equation.
\begin{eqnarray}
2t\cdot w(\loc-\opt)+w(\loc \cap \opt)\geq& 2t\cdot w(\opt-\loc) \nonumber \\
\Rightarrow 2t\cdot w(\loc-\opt)+(2t+1)\cdot w(\loc\cap \opt)\geq& 2t\cdot w(\opt) \nonumber \\
\Rightarrow (2t+1)\cdot w(\loc-\opt)+(2t+1)\cdot w(\loc \cap \opt)\geq& 2t\cdot w(\opt) \nonumber \\
\Rightarrow (2t+1)w(\loc)\geq& 2t\cdot w(\opt) \nonumber \\
\Rightarrow w(\loc)\geq& \frac{2t}{2t+1}\cdot w(\opt)
\end{eqnarray}

\subsubsection{Obtaining a chain decomposition}
We argue about some properties of two-dimensional \mhcp instances,
based on which we get some associations. Consider the optimal solution
which has the maximum number of sets in common with the output of the
local search algorithm.
\begin{enumerate}
\item For each $\ell_i$ we have that $\exists o_j,o_l$ such that $\ell_i\cap OPT\subseteq o_j\cup o_l$. It is simple to see
that if this is not true then we can change the optimal solution so that the number of sets in common with the local optimum increases. Now associate $\ell_i$ to the corresponding $o_j$ and $o_l$. Let $\initialAssoc(o_i)$ be the local sets associated with $o_i$ and $\initialAssoc(\ell_i)$ be the opt sets associated with $\ell_j$.
\item For each $o_i$ we have that $\exists \ell_j,\ell_t\in \initialAssoc(o_i)$ such that $o_i\cap \initialAssoc(o_i)\subseteq \ell_j\cup \ell_t$. This is true due to the different form of local search used. Because otherwise we can change the local optimum to increase the area. Now if $\initialAssoc(o_i)$ has more than two sets $\ell_j$'s. Then among them choose two $\ell_j,\ell_t$ such that $o_i\cap \initialAssoc(o_i)\subseteq \ell_j\cup \ell_t$ and keep the association and remove the rest of the associations for $o_i$. Let the new associations be called $\newAssoc(\ell_i)$ and $\newAssoc(o_j)$.
\item Note that in step 2 we remove some associations. Hence it might no longer be true that $\ell_i\cap OPT\subseteq \newAssoc(\ell_i)$. But also note that it is still true that $o_i\cap \initialAssoc(o_i)\subseteq \newAssoc(o_i)$.
\item Note that due to step 1,2 we have that each local set is associated to at most two opt set and each opt set is associated to at most two local sets.
\item Now form maximal alternating pseudo chains such that a pseudo chain is a list of alternating local and optimal sets. Additionally each local set has the association to its adjacent sets and each opt set has association to its adjacent sets.
\item Now there are three kinds of pseudo chain depending on their end points. They are either $l-l$ or $o-o$ or $o-l$ (here $l-l$ means a chain starting with a local set and ending with a local set).
\item merge $l-l$ pseudo chain arbitrarily with $o-o$ to get only $o-l$ chain.
\end{enumerate}
The $o-l$ chain thus got are the chain we desired in the analysis. It is left to be proven that this decomposition satisfies the properties needed.
\begin{itemize}
\item By construction the local and opt sets alternate (cyclically) in a chain $C_i$.
\item Consider a portion of the chain(cyclically) $...o_1\ell_1o_2...\ell_to_{t+1}...$. Let $L=\{\ell_1,\ell_2,...,\ell_t\}$, $O_1=\{o_1,o_2,...,o_t\}$ and $O_2=\{o_2,o_3,...o_{t+1}\}$. Then we need to prove that $v(L)\cap OPT\subseteq v(L)\cap (O_1\cup O_2)$. The proof is by contradiction. ie. Let $x\in v(L)\cap OPT$ but $x\notin O_1\cup O_2$. Then $x\in \ell_c\in L$. We follow through the associations.
\begin{itemize}
\item Consider the initial association. Then by its property $o_j\in \initialAssoc(\ell_c)$ such that $x\in o_j$. 
\item If $o_j\in \newAssoc(\ell_c)$ then $o_j$ is adjacent to $\ell_c$ in the chain and hence $o_j\in \{o_1,o_2,...,o_{t+1}\}$ which is a contradiction to the fact that $x\notin O_1\cup O_2$.
\item If $o_j\notin \newAssoc(\ell_c)$ then in step 2 of associations $o_j\cap \initialAssoc(o_j)\subseteq \newAssoc(o_j)$. Hence $\ell_d\in \newAssoc(o_j)$ such that $x\in \ell_d$  and $o_j$ and $\ell_d$ are adjacent in some chain. If $\ell_d\in \{\ell_1,\ell_2,...,\ell_t\}$ then $o_j\in \{o_1,o_2,...,o_{t+1}\}$ and we arrive at a contradiction that $x\notin O_1\cup O_2$. Otherwise $x\notin v(L)$ and we still arrive at a contradiction.
\end{itemize}
\end{itemize}

}

\end{document}